\documentclass[reqno]{amsart}

\usepackage{amsmath,amssymb,amsthm,mathtools}
\usepackage{graphicx}
\usepackage{xcolor}
\usepackage[section]{placeins}
\usepackage{booktabs}
\usepackage{subcaption}
\usepackage{url}
\usepackage[colorlinks=true,linkcolor=blue!50!black,citecolor=blue!50!black,urlcolor=blue!50!black]{hyperref}

\newtheorem{theorem}{Theorem}[section]
\newtheorem{proposition}[theorem]{Proposition}
\newtheorem{lemma}[theorem]{Lemma}
\newtheorem{corollary}[theorem]{Corollary}
\theoremstyle{definition}

\newtheorem{example}[theorem]{Example}
\theoremstyle{remark}
\newtheorem{remark}[theorem]{Remark}

\newcommand{\E}{\mathbb{E}}
\newcommand{\Prob}{\mathbb{P}}
\newcommand{\bd}{\partial}
\newcommand{\Pdom}{\Pi_{\mathrm{dom}}}
\newcommand{\Pcul}{\Pi_{\mathrm{cult}}}
\newcommand{\rdom}{r_{\mathrm{dom}}}
\newcommand{\rcul}{r_{\mathrm{cult}}}
\DeclareMathOperator{\vol}{vol}

\title[Absorbing states of the Axelrod model]{Triadic and assortative structure of absorbing states in the Axelrod model}
\author{Moses Boudourides}
\address{School of Professional Studies, Northwestern University, Evanston, IL 60208, USA}
\email{Moses.Boudourides@northwestern.edu}

\begin{document}

\begin{abstract}
We study the Axelrod model of cultural dissemination, viewed as a finite-state Markov chain on an arbitrary finite graph, and classify its absorbing states. An absorbing configuration splits the edges into interior edges, whose endpoints share their whole culture, and boundary edges, whose endpoints share nothing. A partition of the vertices into connected blocks is the domain partition of some absorbing configuration exactly when its quotient graph is properly $q$-colourable. The configurations with a given domain partition correspond to homomorphisms from the quotient into the $F$-fold categorical power of $K_q$, so their total number is a sum of chromatic polynomials over connected partitions; this gives closed forms for trees and cycles and the value $2^{n+F-1}$ for $q=2$ on connected graphs. In absorbing configurations no triangle carries exactly one boundary edge, the open-wedge mass of the interior graph is at most that of the host graph, and Newman's assortativity coefficients for domains, cultures and single features are ordered. The feature counts are martingales and their sums of squares submartingales, which yields an absorption-time bound of order $n^2$ that is attained on paths. Exhaustive enumeration and simulations illustrate the results.
\end{abstract}

\maketitle

\section{Introduction}\label{sec:intro}

Axelrod's model of cultural dissemination \cite{axelrod1997} attaches to each individual a culture, that is, a vector of $F$ features, each taking one of $q$ values. Neighbours interact at a rate proportional to the number of features they share, and an interaction makes them agree on one further feature. The first mechanism is homophily and the second is social influence. Their combination produces a phenomenon absent from the voter model: two neighbours who share nothing never interact again, so the dynamics can freeze in a configuration in which several cultures coexist. The model has been studied extensively by simulation and mean-field methods in statistical physics \cite{castellano2000,vilone2002,klemm2003,castellano2009}, and it remains a reference point for models of social influence \cite{flache2017}. The case $F=q=2$ is closely related to the constrained voter model \cite{vazquez2003,vazquez2004}.

Rigorous results are fewer. Lanchier \cite{lanchier2012} proved that the one-dimensional system with $F=q=2$ clusters and that it fixates in a fragmented configuration when $q$ is sufficiently large relative to $F$; Lanchier and Schweinsberg \cite{lanchier2012schweinsberg} extended clustering to $q=2$ and arbitrary $F$; Lanchier and Scarlatos \cite{lanchier2013scarlatos} proved fixation for $F\le cq$ with $e^{-c}=c$ and for $F=2$, $q\ge3$; and Lanchier and Moisson \cite{lanchier2016moisson} treated features with different numbers of values. A two-dimensional analysis is due to Li \cite{li2014}, and a related model with a confidence threshold was studied in \cite{lanchier2010}. On finite connected graphs, lower bounds on the probability of consensus that do not depend on the topology have been obtained for the multivariate Deffuant model \cite{lanchier2020li} and for imitation and attraction processes with a confidence threshold on a general opinion graph \cite{hardin2022}, the latter building on the framework of \cite{lanchier2017scarlatos}. These works concern the probability of consensus and do not describe the set of absorbing configurations as such. These results concern infinite lattices, mostly the line, where the dynamics of disagreements along edges can be compared with systems of annihilating random walks. Such methods depend on the geometry of $\mathbb{Z}$ and do not extend directly to graphs with many short cycles.

The present paper takes a more modest route and works on an arbitrary finite graph. On a finite graph the process is absorbed almost surely \cite[Lemma~6]{lanchier2012}, and every absorbing configuration has the property that each edge joins two vertices that agree on all features or on none. Our aim is to describe exactly which configurations of this kind exist and what structure they force on the graph. The central result (Theorem~\ref{thm:colour}) is a classification of the cultural domains of absorbing configurations: a partition of the vertices into connected blocks is realised as a domain partition if and only if its quotient graph admits a proper $q$-colouring, and the absorbing configurations with a given domain partition are in bijection with graph homomorphisms from the quotient into the $F$-fold categorical power of the complete graph $K_q$. Counting them gives the number of absorbing configurations as a sum of chromatic polynomials raised to the power $F$ (Corollary~\ref{cor:count}). The term of this sum for the partition into singletons is the number of proper $q$-colourings of $G$, that is, of zero-temperature ground states of the antiferromagnetic $q$-state Potts model on $G$ \cite{sokal2005}, raised to the power $F$; the counting of homomorphisms into complete graphs is classical \cite{birkhoff1912,hell2004,dong2005}. For binary imitation dynamics on finite graphs, Li \cite{li2025} observed that absorbing ``alternating'' states, in which every edge disagrees, exist exactly when the graph is bipartite; this is the special case of Theorem~\ref{thm:colour} with singleton blocks and $q=2$. We have not found the quotient-colouring criterion, or the resulting enumeration, stated for the Axelrod model, although the frozen mixtures of the constrained voter model \cite{vazquez2003,vazquez2004} are absorbing configurations of exactly this type for $F=q=2$. Colourings have been used before in this area: Lanchier \cite{lanchier2010} used the chromatic number to construct absorbing configurations in which many opinions coexist. Moreover, the absorbing configurations of the Axelrod model coincide with those of the imitation process of \cite{hardin2022} on the Hamming opinion graph with threshold $F-1$, and the argument of Theorem~\ref{thm:colour} applies to any threshold model of this kind (Remark~\ref{rem:threshold}).

The other results describe these absorbing configurations from two further angles, using tools developed in earlier work. The first is the two-walk operator $W=A^2-D$ of the adjacency matrix and its unique decomposition into a part supported on edges, which counts triangles, and a part supported on non-edges, which counts open wedges \cite{boudourides_jcn}. The second is the description of the attribute assortativity of a partition through the split of edges into interior and boundary edges \cite{boudourides_dcdsb}. Both apply directly, since an absorbing configuration is an interior--boundary split of the edge set in which the boundary edges carry no shared features at all.

The paper is organised as follows. Section~\ref{sec:martingales} records conservation laws of the dynamics. The feature counts are martingales, which is the familiar voter-model property surviving in the presence of homophily, and the sums of their squares are submartingales; the drift of their total $Q$ is $2/F$ times the total overlap on active edges. The second fact yields the explicit bound $\E\tau\le \tfrac F2\,(Fn^2-Q(\sigma_0))$ on the absorption time, where $Q$ is the total pairwise overlap; this gives a quantitative version of almost sure absorption, whose order in $n$ is attained on paths. The same martingales determine the law of the final state when one feature is initially shared by all vertices. Section~\ref{sec:triadic} shows that in an absorbing configuration no triangle carries exactly one boundary edge, writes this as two Hadamard-product identities for the interior and boundary adjacency matrices, and compares the open-wedge mass of the interior graph with that of the host graph. Section~\ref{sec:quotient} contains the classification and the enumeration of absorbing configurations, with closed forms for trees and cycles, the asymptotics as $q\to\infty$, and a bound on the number of domains in terms of $\chi(G)$. Section~\ref{sec:assortativity} treats the attribute assortativity of absorbing configurations with respect to domains, cultures and single features, and shows that the three coefficients are ordered. Section~\ref{sec:numerics} checks the enumeration exhaustively on small graphs and illustrates the remaining results by simulation, and Section~\ref{sec:discussion} lists open problems.

The individual arguments are elementary, and several of them will look familiar to readers who know the voter model. None of them uses lattice structure, so they apply to every finite graph.

\section{The model and its absorbing configurations}\label{sec:model}

Let $G=(V,E)$ be a finite simple graph with $n=|V|\ge2$ vertices and $m=|E|\ge1$ edges. Unless stated otherwise $G$ is connected, so every vertex has positive degree $d_x$. Write $A$ for the adjacency matrix and $x\sim y$ when $\{x,y\}\in E$. Fix integers $F\ge2$ and $q\ge2$, and let $[q]=\{1,\dots,q\}$. A \emph{configuration} is a map $\sigma\colon V\to[q]^F$; we write $\sigma_f(x)$ for the $f$-th feature of the culture $\sigma(x)$. The \emph{overlap} of two vertices is
\[
o_{xy}(\sigma)=\sum_{f=1}^F \mathbf 1\{\sigma_f(x)=\sigma_f(y)\}\in\{0,1,\dots,F\},
\]
and $h_{xy}=F-o_{xy}$ is the Hamming distance between $\sigma(x)$ and $\sigma(y)$. Since the Hamming distance is a metric on $[q]^F$,
\begin{equation}\label{eq:metric}
o_{xz}\ \ge\ o_{xy}+o_{yz}-F\qquad\text{for all }x,y,z\in V.
\end{equation}

We use the continuous-time formulation of \cite{lanchier2012}. Each edge $\{x,y\}$ is activated at rate one; one of its endpoints, say $x$, is chosen with probability $1/2$; with probability $o_{xy}/F$ the vertex $x$ then copies the value $\sigma_f(y)$ of a feature $f$ chosen uniformly among the $F-o_{xy}$ features on which $x$ and $y$ disagree, and nothing happens otherwise. Let $\sigma^{x\leftarrow y,f}$ denote the configuration obtained from $\sigma$ by setting $\sigma_f(x)=\sigma_f(y)$. The process $(\sigma_t)_{t\ge0}$ is the continuous-time Markov chain on $([q]^F)^V$ with generator
\begin{equation}\label{eq:generator}
\mathcal L\varphi(\sigma)=\sum_{\{x,y\}\in E}\ \sum_{(u,v)\in\{(x,y),(y,x)\}}\ c\bigl(o_{xy}(\sigma)\bigr)\sum_{f:\,\sigma_f(u)\ne\sigma_f(v)}\bigl[\varphi(\sigma^{u\leftarrow v,f})-\varphi(\sigma)\bigr],
\end{equation}
where $c(o)=o/\bigl(2F(F-o)\bigr)$ for $0\le o<F$; when $o_{xy}=F$ the inner sum is empty. The rate $c(o_{xy})$ is symmetric in $x$ and $y$, and this symmetry is behind the conservation laws of Section~\ref{sec:martingales}.

An edge is \emph{interior} if $o_{xy}=F$, \emph{boundary} if $o_{xy}=0$, and \emph{active} otherwise. We write $I(\sigma)$, $\bd(\sigma)$ and $\mathrm{Act}(\sigma)$ for the three classes, so that $E=I\sqcup\bd\sqcup\mathrm{Act}$, and $A_I$, $A_\bd$ for the adjacency matrices of the spanning subgraphs $G_I=(V,I)$ and $G_\bd=(V,\bd)$. The total transition rate out of $\sigma$ equals $\sum_{\{x,y\}\in\mathrm{Act}}o_{xy}/F$, because an active edge with overlap $o$ produces a change with probability $o/F$ per activation. Hence $\sigma$ is absorbing if and only if $\mathrm{Act}(\sigma)=\emptyset$, that is, if and only if every edge is interior or boundary. Let $\mathcal A$ denote the set of absorbing configurations and $\tau=\inf\{t\ge0:\sigma_t\in\mathcal A\}$ the absorption time.

The \emph{cultural domains} of $\sigma$ are the connected components of $G_I$; this is the notion used in \cite{lanchier2012}. They form a partition $\Pdom(\sigma)$ of $V$. We also use the partition $\Pcul(\sigma)$ of $V$ into classes of vertices with the same culture, and, for each feature $f$, the partition $\Pi_f(\sigma)$ into classes of vertices with the same value of feature $f$. Since $\sigma$ is constant along interior edges, $\Pdom$ refines $\Pcul$, and $\Pcul$ refines every $\Pi_f$. Finally, for a configuration $\sigma$ let $N_{f,s}(\sigma)=|\{x\in V:\sigma_f(x)=s\}|$ be the number of vertices holding value $s$ in feature $f$.

\section{Conservation laws and a quadratic Lyapunov function}\label{sec:martingales}

The first statement is the analogue of the classical martingale property of the voter model \cite{liggett1985}; homophily changes the rates but not their symmetry.

\begin{proposition}\label{prop:martingale}
For every feature $f$ and every function $g\colon[q]\to\mathbb R$, the process $\sum_{x\in V}g(\sigma_{t,f}(x))$ is a martingale. In particular each $N_{f,s}(\sigma_t)$ is a martingale and $\E\,N_{f,s}(\sigma_\tau)=\E\,N_{f,s}(\sigma_0)$.
\end{proposition}

\begin{proof}
Let $\varphi(\sigma)=\sum_x g(\sigma_f(x))$. A transition $\sigma\to\sigma^{u\leftarrow v,f'}$ with $f'\ne f$ leaves $\varphi$ unchanged. For an edge $\{x,y\}$ with $\sigma_f(x)\ne\sigma_f(y)$, the two transitions $x\leftarrow y$ and $y\leftarrow x$ in feature $f$ occur at the same rate $c(o_{xy})$ and change $\varphi$ by $g(\sigma_f(y))-g(\sigma_f(x))$ and $g(\sigma_f(x))-g(\sigma_f(y))$ respectively. Their contributions to \eqref{eq:generator} cancel, so $\mathcal L\varphi\equiv0$. On a finite state space this makes $\varphi(\sigma_t)$ a martingale. The process is bounded and $\tau<\infty$ almost surely (Corollary~\ref{cor:time} below, or \cite[Lemma~6]{lanchier2012}), so optional stopping gives the last identity.
\end{proof}

Next consider, for each feature $f$, the quantity
\begin{align*}
Q_f(\sigma)&=\sum_{s\in[q]}N_{f,s}(\sigma)^2=\bigl|\{(x,y)\in V^2:\sigma_f(x)=\sigma_f(y)\}\bigr|,\\
Q(\sigma)&=\sum_{f=1}^FQ_f(\sigma)=\sum_{(x,y)\in V^2}o_{xy}(\sigma),
\end{align*}
where the sums run over ordered pairs, diagonal included. Thus $Q$ is the total pairwise overlap of the population, taken over all pairs and not only over edges, and $Q_f/n^2$ is the Simpson concentration of feature $f$. We call $Q$ a Lyapunov function in the stochastic sense: it need not increase along every sample path, but its generator drift is nonnegative and, as shown next, bounded away from zero before absorption.

\begin{theorem}\label{thm:lyapunov}
For every feature $f$,
\begin{equation}\label{eq:LQf}
\mathcal LQ_f(\sigma)=\frac2F\sum_{\substack{\{x,y\}\in E\\ \sigma_f(x)\ne\sigma_f(y)}}\frac{o_{xy}}{F-o_{xy}}\ \ge 0,
\qquad\text{and}\qquad
\mathcal LQ(\sigma)=\frac2F\sum_{\{x,y\}\in\mathrm{Act}(\sigma)}o_{xy}.
\end{equation}
Consequently each $Q_f(\sigma_t)$ and $Q(\sigma_t)$ is a bounded submartingale, and
\begin{equation}\label{eq:Qidentity}
\E\,Q(\sigma_\tau)-\E\,Q(\sigma_0)=\frac2F\,\E\int_0^\tau\sum_{\{x,y\}\in\mathrm{Act}(\sigma_t)}o_{xy}(\sigma_t)\,dt .
\end{equation}
\end{theorem}

\begin{proof}
Suppose $\sigma_f(u)=a\ne b=\sigma_f(v)$ and $u$ copies feature $f$ from $v$. Then $N_{f,a}$ decreases by one and $N_{f,b}$ increases by one, so $Q_f$ changes by $(N_{f,b}+1)^2-N_{f,b}^2+(N_{f,a}-1)^2-N_{f,a}^2=2(N_{f,b}-N_{f,a})+2$. The reverse transition, in which $v$ copies $a$ from $u$, changes $Q_f$ by $2(N_{f,a}-N_{f,b})+2$. Both occur at rate $c(o_{uv})$, so the edge contributes $4c(o_{uv})=2o_{uv}/\bigl(F(F-o_{uv})\bigr)$ to $\mathcal LQ_f$. Summing over edges that disagree in $f$ gives the first formula; the terms with $o_{xy}=0$ vanish. Summing over $f$, an active edge disagrees in exactly $F-o_{xy}$ features, so it contributes $(F-o_{xy})\cdot 2o_{xy}/(F(F-o_{xy}))=2o_{xy}/F$ to $\mathcal LQ$, while boundary and interior edges contribute nothing. For the identity, Dynkin's formula for the bounded stopping time $\tau\wedge t$ gives $\E Q(\sigma_{\tau\wedge t})-\E Q(\sigma_0)=\E\int_0^{\tau\wedge t}\mathcal LQ(\sigma_s)\,ds$. As $t\to\infty$, the left side converges by bounded convergence, since $\tau<\infty$ almost surely by Corollary~\ref{cor:time}, whose proof uses only this display for finite $t$, and the right side converges by monotone convergence.
\end{proof}

\begin{corollary}\label{cor:time}
For every initial configuration $\sigma_0$,
\[
\E\,\tau\ \le\ \frac F2\bigl(Fn^2-Q(\sigma_0)\bigr)\ \le\ \frac{F^2n(n-1)}2 .
\]
In particular the process is absorbed almost surely.
\end{corollary}

\begin{proof}
If $\sigma\notin\mathcal A$, at least one edge is active and its overlap is at least one, so $\mathcal LQ(\sigma)\ge 2/F$ by \eqref{eq:LQf}. Dynkin's formula for $\tau\wedge t$ therefore gives
$\frac2F\,\E[\tau\wedge t]\le \E Q(\sigma_{\tau\wedge t})-Q(\sigma_0)\le Fn^2-Q(\sigma_0)$, because $Q\le Fn^2$. Letting $t\to\infty$ and using monotone convergence yields the first inequality. The second follows from $Q(\sigma_0)\ge Fn$, the contribution of the diagonal pairs.
\end{proof}

\begin{remark}
Almost sure absorption on finite graphs is Lemma~6 of \cite{lanchier2012}; Corollary~\ref{cor:time} adds an explicit bound of order $F^2n^2$, uniform over all graphs on $n$ vertices. The argument is the familiar quadratic-Lyapunov argument for voter-type models, with the homophily factor $o_{xy}/F$ absorbed into the drift. The bound is crude; in the simulations of Section~\ref{sec:numerics} the mean absorption times are one to three orders of magnitude smaller. Proposition~\ref{prop:sharp} below shows that its order in $n$ is nevertheless attained on paths.
\end{remark}

\begin{corollary}\label{cor:concentration}
$\E\,Q_f(\sigma_\tau)\ge\E\,Q_f(\sigma_0)$ for every $f$, and $\E\,Q(\sigma_\tau)\ge\E\,Q(\sigma_0)+\frac2F\,\E\tau$.
\end{corollary}

\begin{proof}
The first statement is the submartingale property with optional stopping. The second follows from \eqref{eq:Qidentity} together with $\sum_{\mathrm{Act}}o_{xy}\ge1$ before absorption.
\end{proof}

In words, the Simpson concentration of every feature can only increase in expectation, and the total overlap gains at least $2/F$ per unit of time spent before absorption. Section~\ref{sec:cliques} turns this into a statement about domain sizes on complete graphs.

The order $n^2$ in Corollary~\ref{cor:time} cannot be improved in general, while the factor $F^2$ can.

\begin{proposition}\label{prop:sharp}
\textup{(i)} On the path $P_n$ with vertices $0,\dots,n-1$, let $\sigma_0$ equal a culture $c$ on $\{0,\dots,j-1\}$ and a culture $c'$ on $\{j,\dots,n-1\}$, where $c$ and $c'$ differ in exactly one feature. Then $\E\tau=j(n-j)F/(F-1)$; in particular $\sup_{\sigma_0}\E\tau\ge\lfloor n^2/4\rfloor$.
\textup{(ii)} On $K_2$ started from overlap $o_0\ge1$, $\E\tau=F\,(H_{F-1}-H_{o_0-1})$, where $H_k=\sum_{i=1}^k1/i$; the largest value is $F H_{F-1}$, of order $F\log F$.
\end{proposition}

\begin{proof}
(i) At every time before absorption the configuration consists of two blocks of cultures $c$ and $c'$, and the only active edge is the one joining them, with overlap $F-1$. It fires at rate $(F-1)/F$, and each firing moves the interface one step to the left or to the right with probability $1/2$. The position of the interface is therefore a simple random walk run at rate $(F-1)/F$ and absorbed at $0$ and $n$, and the expected number of steps from $j$ is $j(n-j)$. (ii) The overlap of the two vertices never decreases; from overlap $o\in\{1,\dots,F-1\}$ it waits an exponential time with rate $o/F$ and then increases by one. Summing the expected waiting times gives the formula.
\end{proof}

When some feature is initially shared by everybody, the law of the final state is determined exactly.

\begin{proposition}\label{prop:selection}
Let $G$ be connected, and suppose that some feature $f_0$ is constant in $\sigma_0$. Then $\sigma_\tau$ is a consensus almost surely, and $\Prob(\sigma_{\tau,f}\equiv s)=\E N_{f,s}(\sigma_0)/n$ for every feature $f$ and value $s$.
\end{proposition}

\begin{proof}
A copy reproduces a value already present at the same feature, so $f_0$ stays constant for all times. Every pair of vertices then has overlap at least one, no boundary edge can appear, and an absorbing configuration on a connected graph without boundary edges is a consensus. At consensus $N_{f,s}(\sigma_\tau)=n\,\mathbf 1\{\sigma_{\tau,f}\equiv s\}$, and Proposition~\ref{prop:martingale} gives the law.
\end{proof}

Without such a shared feature the probability of reaching a particular domain partition is not determined by the martingales alone. On $K_2$, for instance, the final state has two domains if and only if $o_0=0$, an event of probability $((q-1)/q)^F$ under uniform initial cultures.

\section{Triadic structure of absorbing configurations}\label{sec:triadic}

Throughout this section $\sigma\in\mathcal A$ is absorbing, so $E=I\sqcup\bd$ and $A=A_I+A_\bd$.

\begin{lemma}\label{lem:induced}
Let $D$ be a domain of $\sigma$. Then $\sigma$ is constant on $D$, and every edge of $G$ with both endpoints in $D$ is interior. Hence $G_I$ is the disjoint union of the induced subgraphs $G[D]$, $D\in\Pdom$, and $\bd$ is the set of edges joining distinct domains.
\end{lemma}

\begin{proof}
The culture is constant along interior edges and $D$ is connected in $G_I$, so $\sigma$ is constant on $D$. If $x,y\in D$ and $x\sim y$, then $o_{xy}=F$ and the edge is interior. An edge joining two distinct domains cannot be interior, since its endpoints would lie in the same component of $G_I$; being an edge of an absorbing configuration, it is a boundary edge.
\end{proof}

The next result is the triadic constraint. In the notation of \cite{boudourides_jcn}, let $t_{\{x,z\}}=(A^2)_{xz}$ denote the number of triangles through the edge $\{x,z\}$, and let $\circ$ denote the Hadamard product.

\begin{theorem}\label{thm:triadic}
Let $\sigma\in\mathcal A$. Then
\begin{equation}\label{eq:hadamard}
(A_I^{\,2})\circ A_\bd=0\qquad\text{and}\qquad(A_IA_\bd)\circ A_I=0 .
\end{equation}
Equivalently, every triangle of $G$ contains $0$, $2$ or $3$ boundary edges. Moreover:
\begin{enumerate}
\item[\textup{(i)}] the triangles through an interior edge have their two remaining edges both interior or both boundary, and the triangles through a boundary edge contain at least one further boundary edge;
\item[\textup{(ii)}] a triangle with three boundary edges carries three pairwise feature-disjoint cultures, so such triangles exist only if $q\ge3$.
\end{enumerate}
\end{theorem}

\begin{proof}
The entry $\bigl((A_I^2)\circ A_\bd\bigr)_{xz}$ counts the vertices $y$ with $\{x,y\},\{y,z\}\in I$ and $\{x,z\}\in\bd$. For such $y$ the vertices $x,y,z$ lie in one domain, and Lemma~\ref{lem:induced} forces $\{x,z\}\in I$, a contradiction. Likewise $\bigl((A_IA_\bd)\circ A_I\bigr)_{xz}$ counts $y$ with $\{x,y\}\in I$, $\{y,z\}\in\bd$ and $\{x,z\}\in I$; then $x,y,z$ lie in one domain and $\{y,z\}\in I$, again a contradiction. Every nonzero contribution to either Hadamard product corresponds to an ordered placement of a path of length two inside a triangle with exactly one boundary edge, and every such triangle, consisting of two interior edges sharing a vertex and a boundary edge closing them, contributes to the first product. This proves the equivalence, and (i) is a restatement. For (ii), the three cultures pairwise disagree in every feature, so feature $1$ takes three distinct values on the triangle.
\end{proof}

The constraint is a property of any partition of $V$ in which $I$ is taken to be the set of edges inside blocks. What is specific to the Axelrod model is that absorbing configurations realise such splits with boundary edges whose endpoints share no feature, which is what Section~\ref{sec:quotient} exploits.

Part (i) refines the triangle multiplicity of \cite{boudourides_jcn} in the absorbing state: writing $t_e=t_e^{(0)}+t_e^{(2)}+t_e^{(3)}$ according to the number of boundary edges of the triangle, an interior edge has $t_e^{(3)}=0$ and a boundary edge has $t_e^{(0)}=0$. The identities \eqref{eq:hadamard} are the absorbing case of a constraint that holds at all times. By \eqref{eq:metric}, for every edge $\{x,z\}$ with $t_{\{x,z\}}\ge1$,
\begin{equation}\label{eq:protection}
o_{xz}\ \ge\ \frac1{t_{\{x,z\}}}\sum_{y\sim x,\ y\sim z}\bigl(o_{xy}+o_{yz}\bigr)-F,
\end{equation}
so an edge whose triangles have, on average, legs with large overlap cannot currently be a boundary edge. In particular an edge closing a wedge with $o_{xy}+o_{yz}>F$ is never a boundary edge.

A wedge of a graph $H$ is an unordered pair of edges sharing a vertex, its middle vertex; it is open if the other two endpoints are not adjacent. Recall from \cite{boudourides_jcn} that the open-wedge mass $\omega(H)=\sum_x\binom{d_x}2-3\Delta(H)$ of a graph $H$ with $\Delta(H)$ triangles (here $\Delta$ never denotes the maximum degree) is the number of open wedges of $H$, each counted once at its middle vertex, and that $\omega(H)=0$ exactly when $H$ is a disjoint union of cliques.

\begin{corollary}\label{cor:omega}
Let $\sigma\in\mathcal A$. Then $\omega(G_I)=\sum_{D\in\Pdom}\omega(G[D])$ and
\[
\omega(G)-\omega(G_I)=\bigl|\{\text{open wedges of }G\text{ with at least one boundary leg}\}\bigr|\ \ge0 .
\]
In particular $\omega(G_I)=\omega(G)$ if and only if every open wedge of $G$ has both legs interior. Likewise $\Delta(G)=\sum_{D}\Delta(G[D])+\Delta_2+\Delta_3$, where $\Delta_k$ is the number of triangles with exactly $k$ boundary edges.
\end{corollary}

\begin{proof}
The first identity holds because $G_I$ is the disjoint union of the induced subgraphs $G[D]$ (Lemma~\ref{lem:induced}). A wedge of $G_I$ is a wedge $(x,y,z)$ of $G$ with both legs interior; its three vertices lie in one domain, so by Lemma~\ref{lem:induced} the pair $\{x,z\}$ is an edge of $G_I$ if and only if it is an edge of $G$. The wedge is therefore open in $G_I$ exactly when it is open in $G$. Hence the open wedges of $G_I$ are the open wedges of $G$ with two interior legs, and the difference counts those with a boundary leg. The triangle identity follows from Theorem~\ref{thm:triadic}, since the triangles with no boundary edge are exactly the triangles of the graphs $G[D]$.
\end{proof}

\section{Domain quotients and colourability}\label{sec:quotient}

For a partition $\Pi$ of $V$ into blocks inducing connected subgraphs (a \emph{connected partition}), let $G/\Pi$ denote the simple quotient graph whose vertices are the blocks, two blocks being adjacent when some edge of $G$ joins them. Write $\chi(H)$ for the chromatic number of a graph $H$ and $P(H;q)$ for its chromatic polynomial, the number of proper colourings of $H$ with $q$ colours \cite{birkhoff1912,dong2005}.

\begin{theorem}\label{thm:colour}
A connected partition $\Pi$ of $V$ is the domain partition of some absorbing configuration if and only if $\chi(G/\Pi)\le q$. More precisely, the absorbing configurations with domain partition $\Pi$ are in bijection with $F$-tuples of proper $q$-colourings of $G/\Pi$; there are $P(G/\Pi;q)^F$ of them.
\end{theorem}

\begin{proof}
Let $\sigma\in\mathcal A$ with $\Pdom(\sigma)=\Pi$. By Lemma~\ref{lem:induced}, $\sigma$ is constant on blocks, and every edge between distinct blocks is a boundary edge. Hence for each $f$ the map sending a block to the common value of $\sigma_f$ on it is a proper $q$-colouring $\kappa_f$ of $G/\Pi$, and $\sigma$ is recovered from $(\kappa_1,\dots,\kappa_F)$. Conversely, let $(\kappa_1,\dots,\kappa_F)$ be proper $q$-colourings of $G/\Pi$ and set $\sigma_f(x)=\kappa_f(P)$ for $x$ in block $P$. Edges inside a block have overlap $F$. An edge between adjacent blocks $P\ne P'$ has overlap $0$, because $\kappa_f(P)\ne\kappa_f(P')$ for every $f$. So $\sigma$ is absorbing, its interior edges are exactly the edges inside blocks, and, the blocks being connected, its domains are exactly the blocks. The two constructions are mutually inverse. A proper $q$-colouring exists if and only if $\chi(G/\Pi)\le q$, and its repetition in all $F$ features gives existence.
\end{proof}

\begin{remark}\label{rem:hom}
Theorem~\ref{thm:colour} has a compact formulation in terms of graph homomorphisms \cite{hell2004}. Let $K_q^{\times F}$ denote the $F$-fold categorical (tensor) power of $K_q$: its vertices are the cultures $[q]^F$, and two cultures are adjacent when they differ in every feature. A proper $q$-colouring of a graph $H$ is a homomorphism $H\to K_q$, so that $P(H;q)=\hom(H,K_q)$, and a homomorphism into a categorical product is a tuple of homomorphisms into the factors. The theorem therefore says that the absorbing configurations with domain partition $\Pi$ are exactly the maps $V\to[q]^F$ that are constant on blocks and induce a homomorphism $G/\Pi\to K_q^{\times F}$, and that their number is $\hom(G/\Pi,K_q^{\times F})=\hom(G/\Pi,K_q)^F$. For the partition into singletons, $G/\Pi=G$ and the absorbing configurations in which every edge is a boundary edge are the homomorphisms $G\to K_q^{\times F}$; they exist if and only if $\chi(G)\le q$. With $q=2$ this recovers the observation of Li \cite{li2025} that ``alternating'' absorbing states of binary imitation dynamics exist exactly on bipartite graphs, and, when $\chi(G)\le q$, the count $P(G;q)$ is the number of zero-temperature ground states of the antiferromagnetic $q$-state Potts model on $G$ \cite{sokal2005}.
\end{remark}

\begin{remark}\label{rem:threshold}
Nothing in the proof of Theorem~\ref{thm:colour} uses the specific Axelrod rates; it uses only the shape of the absorbing set. Consider any process in which neighbours with opinions in a finite set $\mathcal O$ interact only if their opinions are distinct and compatible, and stop interacting once they agree, as in the voter models with confidence threshold on an opinion graph $\Gamma$ studied in \cite{lanchier2017scarlatos,hardin2022}, where $a,b\in\mathcal O$ are compatible when $0<d_\Gamma(a,b)\le\tau$. A configuration is then absorbing if and only if every edge joins equal or incompatible opinions. Let $\mathcal I$ be the incompatibility graph on $\mathcal O$, in which $a\ne b$ are adjacent when $d_\Gamma(a,b)>\tau$. The proof of Theorem~\ref{thm:colour} then shows, word for word, that the absorbing configurations with domain partition $\Pi$ correspond to homomorphisms $G/\Pi\to\mathcal I$, and that there are $\sum_\Pi\hom(G/\Pi,\mathcal I)$ absorbing configurations in all. For the Axelrod model, $\Gamma$ is the Hamming graph on $[q]^F$, $\tau=F-1$, and $\mathcal I=K_q^{\times F}$. For opinions in $[0,1]$ with a confidence threshold $\varepsilon$, the singleton term is the construction behind the bounds of \cite{lanchier2010} on the number of opinions that can coexist at equilibrium.
\end{remark}

\begin{corollary}\label{cor:count}
The number of absorbing configurations of the Axelrod model on $G$ is
\[
|\mathcal A|=\sum_{\Pi}P(G/\Pi;q)^F,
\]
the sum running over all connected partitions of $V$; the terms with $\chi(G/\Pi)>q$ vanish.
\end{corollary}

\begin{example}
For $G=K_2$ there are two connected partitions, with quotients $K_1$ and $K_2$, so $|\mathcal A|=q^F+\bigl(q(q-1)\bigr)^F$. These are the pairs of equal cultures and the pairs of cultures differing in every feature.
\end{example}

For two values per feature the sum in Corollary~\ref{cor:count} collapses to a quantity that does not depend on the graph.

\begin{corollary}\label{cor:q2}
Let $G$ be connected and $q=2$. Then $|\mathcal A|=2^{\,n+F-1}$, and exactly $2^{n-1}$ connected partitions of $V$ have a bipartite quotient.
\end{corollary}

\begin{proof}
When $q=2$, each culture $c$ has a unique feature-disjoint partner, its complement $\bar c$. A configuration is therefore absorbing if and only if every edge joins equal or complementary cultures. Fix a vertex $x_0$. For each culture $c$ and each set $S\subseteq V\setminus\{x_0\}$, the configuration equal to $\bar c$ on $S$ and to $c$ elsewhere is absorbing, and every absorbing configuration arises in this way from exactly one pair $(c,S)$, because connectedness forces every culture to lie in $\{\sigma(x_0),\overline{\sigma(x_0)}\}$. This gives $2^F\cdot2^{n-1}$ configurations. On the other hand, a connected bipartite graph has exactly two proper $2$-colourings, so each connected partition with bipartite quotient contributes $2^F$ to the sum in Corollary~\ref{cor:count}, and the other partitions contribute nothing. Comparing the two counts gives the second statement.
\end{proof}

Since $\mathcal A$ factorises over connected components, a graph with $c$ components has $|\mathcal A|=2^{\,n+c(F-1)}$ absorbing configurations when $q=2$.

The whole count can also be written as a single homomorphism number, which gives closed forms for trees and cycles. Let $H_{q,F}$ be the graph $K_q^{\times F}$ of Remark~\ref{rem:hom} with a loop added at every vertex; its adjacency matrix is $M=I+(J_q-I_q)^{\otimes F}$, where $J_q$ is the all-ones $q\times q$ matrix.

\begin{proposition}\label{prop:closed}
For every graph $G$, $|\mathcal A|=\hom(G,H_{q,F})$. Consequently, for a tree $T$ on $n$ vertices and for the cycle $C_n$,
\[
|\mathcal A(T)|=q^F\bigl(1+(q-1)^F\bigr)^{n-1},
\]
\[
|\mathcal A(C_n)|=\operatorname{tr}M^n=\sum_{j=0}^F\binom Fj(q-1)^j\bigl(1+(-1)^j(q-1)^{F-j}\bigr)^n .
\]
\end{proposition}

\begin{proof}
A configuration is absorbing exactly when each edge joins equal cultures, which corresponds to a loop of $H_{q,F}$, or feature-disjoint cultures, which corresponds to an edge of $K_q^{\times F}$; this is the first identity, and sorting the homomorphisms according to the set of edges sent to loops recovers Corollary~\ref{cor:count}. For a tree, choose a root culture in $q^F$ ways; each further vertex, taken in breadth-first order, either copies its parent's culture or takes one of the $(q-1)^F$ cultures feature-disjoint from it. For the cycle, $\hom(C_n,H)=\operatorname{tr}M^n$. The matrix $J_q-I_q$ has eigenvalues $q-1$ (once) and $-1$ ($q-1$ times), and eigenvalues of Kronecker products are products of eigenvalues. Hence $M$ has eigenvalue $1+(-1)^j(q-1)^{F-j}$ with multiplicity $\binom Fj(q-1)^j$, for $j=0,\dots,F$.
\end{proof}

With $F=1$ the first identity reduces to the classical $\sum_\Pi P(G/\Pi;q)=q^n$, since every configuration of a one-feature model is absorbing. In the opposite direction, the proportion of absorbing configurations tends to one as $q$ grows. Indeed $P(G;q)=q^n-mq^{n-1}+O(q^{n-2})$, and every other connected partition has at most $n-1$ blocks and contributes $O(q^{(n-1)F})$, so that for $F\ge2$
\[
\frac{|\mathcal A|}{q^{nF}}=1-\frac{Fm}{q}+O(q^{-2})\qquad(q\to\infty).
\]
This is the finite-graph counterpart of the fragmentation observed for large $q$ \cite{castellano2000,lanchier2012}: almost every configuration is already frozen.

\begin{corollary}\label{cor:cliques}
Let $\sigma\in\mathcal A$.
\begin{enumerate}
\item[\textup{(i)}] Every clique of $G$ meets at most $q$ domains.
\item[\textup{(ii)}] If $q=2$, the quotient $G/\Pdom$ is bipartite. In particular it has no triangles, and every triangle of $G$ has $0$ or $2$ boundary edges.
\item[\textup{(iii)}] (Here $G$ need not be connected.) If $G$ is a disjoint union of cliques, equivalently $\omega(G)=0$, each component of $G$ carries at most $q$ domains, and their cultures are pairwise feature-disjoint.
\end{enumerate}
\end{corollary}

\begin{proof}
The vertices of a clique lying in $k$ distinct domains give $k$ pairwise adjacent blocks of $G/\Pdom$, so $k\le\chi(G/\Pdom)\le q$ by Theorem~\ref{thm:colour}; this is (i), and (iii) is (i) applied to each component together with Lemma~\ref{lem:induced}. For (ii), a $2$-colourable graph is bipartite, and a triangle of $G$ with three boundary edges would meet three domains forming a triangle of the quotient.
\end{proof}

The chromatic number of the host graph also bounds the number of domains.

\begin{corollary}\label{cor:chi}
Every absorbing configuration has at most $n-\chi(G)+q$ domains and at least $\chi(G)-q$ interior edges.
\end{corollary}

\begin{proof}
Contracting an edge lowers the chromatic number by at most one, since a colouring of the contracted graph extends to the original graph after one endpoint receives a new colour. The quotient $G/\Pdom$ is obtained from $G$ by contracting the $n-|\Pdom|$ edges of spanning trees of the domains (and deleting loops and parallel edges), so $\chi(G)\le\chi(G/\Pdom)+n-|\Pdom|\le q+n-|\Pdom|$ by Theorem~\ref{thm:colour}. Each domain is connected in $G_I$, hence $|I|\ge n-|\Pdom|\ge\chi(G)-q$.
\end{proof}

For $G=K_n$ the bound reads $|\Pdom|\le q$ and is attained; in general it is informative only when $\chi(G)>q$.

\begin{remark}
The domain quotient is the natural setting for the contraction inequality of \cite[Theorem~10.6]{boudourides_jcn}. With $B$ the directed edge-sum quotient matrix of $\Pdom$ and $M$ the aggregated two-walk matrix, $B_{aa}=2|E(G[D_a])|$ counts interior edge-ends, the off-diagonal entries $B_{ab}$ count boundary edges, and $M\le B^2$ entrywise, with an explicit overcount. Theorem~\ref{thm:colour} adds that the cultures of $B$-adjacent blocks are feature-disjoint. A closer study of $M$ and $B^2$ in absorbing states, for instance through the wedge-equitability of $\Pdom$, is left for future work.
\end{remark}

\section{Assortativity of absorbing configurations}\label{sec:assortativity}

Let $\Pi$ be a partition of $V$. Following Newman \cite{newman2003mixing}, let $e_{\mathrm{in}}(\Pi)$ be the fraction of edges with both endpoints in the same block, let $s(\Pi)=\sum_{P\in\Pi}\bigl(\vol P/2m\bigr)^2$ with $\vol P=\sum_{x\in P}d_x$, and, when $s(\Pi)<1$, let
\[
r(\Pi)=\frac{e_{\mathrm{in}}(\Pi)-s(\Pi)}{1-s(\Pi)}
\]
be the attribute assortativity coefficient of the labelling of vertices by blocks. In the language of \cite{boudourides_dcdsb}, $e_{\mathrm{in}}$ is the interior share of the edge set relative to $\Pi$ and $1-e_{\mathrm{in}}$ its boundary share; the coefficient compares the interior share with its value under degree-preserving mixing. For an absorbing configuration we abbreviate $\rdom=r(\Pdom)$, $\rcul=r(\Pcul)$ and $r_f=r(\Pi_f)$, and we write $\iota=|I|/m$ for the interior fraction.

\begin{theorem}\label{thm:chain}
Let $G$ be connected and let $\sigma\in\mathcal A$ not be a consensus. Then:
\begin{enumerate}
\item[\textup{(i)}] $e_{\mathrm{in}}(\Pdom)=e_{\mathrm{in}}(\Pcul)=e_{\mathrm{in}}(\Pi_f)=\iota<1$ for every $f$;
\item[\textup{(ii)}] $s(\Pdom)\le s(\Pcul)\le s(\Pi_f)<1$ for every $f$;
\item[\textup{(iii)}] $r_f\le\rcul\le\rdom$ for every $f$. Equality $\rcul=\rdom$ holds if and only if no two domains carry the same culture, and $r_f=\rcul$ if and only if $\Pi_f=\Pcul$;
\item[\textup{(iv)}] if $\Pi\in\{\Pdom,\Pcul,\Pi_f\}$ has $k$ blocks, then $r(\Pi)\le(k\iota-1)/(k-1)$, with equality if and only if the $k$ blocks have equal volume; in particular $r_f\le(q\iota-1)/(q-1)$.
\end{enumerate}
\end{theorem}

\begin{proof}
(i) Since $\Pdom$ refines $\Pcul$ and $\Pcul$ refines $\Pi_f$, an edge inside a block of a finer partition lies inside a block of each coarser one. By Lemma~\ref{lem:induced} the edges inside blocks of $\Pdom$ are exactly the interior edges. A boundary edge joins two cultures that differ in every feature, so it does not lie inside a block of $\Pi_f$. Hence all three partitions have the interior edges as their internal edges. Because $\sigma$ is not a consensus and $G$ is connected, $G_I\ne G$, so $\bd\neq\emptyset$ and $\iota<1$.

(ii) Each block has positive volume because $G$ is connected and $n\ge2$. Merging two blocks with volumes $a,b>0$ replaces $a^2+b^2$ by $(a+b)^2>a^2+b^2$, so $s$ strictly increases under every proper coarsening. Moreover $\Pi_f$ has at least two blocks, since the endpoints of a boundary edge differ in feature $f$, hence $s(\Pi_f)<1$.

(iii) By (i) the three coefficients have the form $\phi(s)=(\iota-s)/(1-s)$ with a common $\iota<1$, and $\phi'(s)=(\iota-1)/(1-s)^2<0$ on $[0,1)$. Combining with (ii) gives the chain. Since $\phi$ is strictly decreasing and $s$ strictly increases under proper coarsening, equality holds exactly when the two partitions coincide. Now $\Pcul=\Pdom$ means that distinct domains carry distinct cultures.

(iv) By the Cauchy--Schwarz inequality, $s(\Pi)\ge1/k$, and $\phi$ is decreasing, so $r(\Pi)\le\phi(1/k)=(k\iota-1)/(k-1)$. The right side is increasing in $k$, its derivative being $(1-\iota)/(k-1)^2>0$, and $\Pi_f$ has at most $q$ blocks. Equality in $s(\Pi)\ge1/k$, and hence in the bound for $r(\Pi)$, holds exactly when all blocks have volume $2m/k$.
\end{proof}

For an arbitrary pair of nested partitions the finer one never has the larger internal-edge fraction, but it may have a strictly smaller one, and the ordering of the assortativity coefficients can then go either way. In an absorbing state the boundary edges disagree in every feature, so the interior fraction is the same at every level of description. All differences between $r_f$, $\rcul$ and $\rdom$ then come from the degree-weighted concentration $s$, that is, from the coexistence of disconnected domains with equal cultures (for $\rcul<\rdom$) and from distinct cultures sharing a feature value (for $r_f<\rcul$).

\subsection{Complete graphs and unions of cliques}\label{sec:cliques}

On a clique the structure of absorbing states is rigid, and the submartingales of Section~\ref{sec:martingales} translate into a statement about domain sizes.

\begin{proposition}\label{prop:clique}
Let $G=K_n$ with $n\ge2$ vertices, so that $m=n(n-1)/2$, and let $\sigma\in\mathcal A$ have $k$ domains $D_1,\dots,D_k$.
\begin{enumerate}
\item[\textup{(i)}] $k\le\min(q,n)$, the cultures of distinct domains are pairwise feature-disjoint, and $\Pdom=\Pcul=\Pi_f$ for every $f$. If $k\ge2$, then $\rdom=-1/(n-1)$, whatever the domain sizes.
\item[\textup{(ii)}] For the Axelrod process on $K_n$ started from $\sigma_0$,
\[
\E\sum_{i}|D_i(\sigma_\tau)|^2\ \ge\ \max\Bigl\{\max_f Q_f(\sigma_0),\ \frac{Q(\sigma_0)}F+\frac2{F^2}\,\E\tau\Bigr\}.
\]
\end{enumerate}
If $G$ is a disjoint union of cliques, part (i) applies to each component separately, and part (ii) applies to the process restricted to each component, with that component's own absorption time.
\end{proposition}

\begin{proof}
(i) Every two domains are adjacent in $K_n$, so the quotient is $K_k$ and Theorem~\ref{thm:colour} gives $k\le q$ and pairwise feature-disjoint cultures. Hence the value of each feature determines the domain, and the three partitions coincide. For block sizes $n_1,\dots,n_k$ and $S=\sum_in_i^2/n^2$, the volumes are $n_i(n-1)$ and $2m=n(n-1)$, so $s=S$, and $e_{\mathrm{in}}=\sum_in_i(n_i-1)/\bigl(n(n-1)\bigr)=(nS-1)/(n-1)$. Therefore $r=\bigl((nS-1)/(n-1)-S\bigr)/(1-S)=-1/(n-1)$.

(ii) By (i), at absorption the classes of every feature are the domains, so $Q_f(\sigma_\tau)=\sum_i|D_i|^2$ for every $f$ and $Q(\sigma_\tau)=F\sum_i|D_i|^2$. The bound follows from Corollary~\ref{cor:concentration}. For a disjoint union of cliques, the coordinates on different components evolve as independent Axelrod processes, because every transition involves an edge inside one component. The argument therefore applies to each component $C$, with its own absorption time $\tau_C$; the absorption time of the whole process is $\max_C\tau_C$.
\end{proof}

Part (i) shows that the assortativity coefficient carries no information on a clique: every non-trivial partition of $K_n$ has the same value $-1/(n-1)$. The orderings of Theorem~\ref{thm:chain} can therefore be strict only on connected graphs that are not complete. Part (ii) is a quantitative form of the tendency towards few large domains. The expected Simpson concentration $\sum_i|D_i|^2/n^2$ of the final domains is at least the largest initial feature concentration, and it exceeds the mean initial overlap by an amount proportional to the expected absorption time.

\section{Exact checks and numerical illustration}\label{sec:numerics}

\subsection{Exhaustive enumeration and validation of the simulator}

Corollary~\ref{cor:count} can be checked directly on small graphs. For nine graphs on three to five vertices we enumerated all $q^{Fn}$ configurations, counted those in which every edge has overlap $0$ or $F$, and compared the count with $\sum_\Pi P(G/\Pi;q)^F$, computed by listing the connected partitions and counting the proper colourings of each quotient by brute force. The two numbers coincided in all $36$ cases of Table~\ref{tab:enum}. The column $q=2$ shows the value $2^{n+F-1}$ of Corollary~\ref{cor:q2}, independent of the graph, A configuration that is absorbing on $G+e$ is absorbing on $G$, so the count can only decrease when an edge is added; for $q=3$ it falls from the two trees through $C_4$ and the paw to the diamond and $K_4$. The values for trees and cycles agree with the closed forms of Proposition~\ref{prop:closed}.

\begin{table}[htbp]
\centering\small
\caption{Number of absorbing configurations $|\mathcal A|$ on small graphs, obtained by exhaustive enumeration of all $q^{Fn}$ configurations; in every case the count equals $\sum_\Pi P(G/\Pi;q)^F$ of Corollary~\ref{cor:count}. The paw is a triangle with a pendant edge, the diamond is $K_4$ minus an edge, and the bull is a triangle with two pendant edges at distinct vertices.}\label{tab:enum}
\begin{tabular}{lrrrrr}
\toprule
Graph & $n$ & $(F,q)=(2,2)$ & $(2,3)$ & $(3,2)$ & $(3,3)$\\
\midrule
$K_3$ & 3 & 16 & 153 & 32 & 891 \\
$P_4$ & 4 & 32 & 1{,}125 & 64 & 19{,}683 \\
$K_{1,3}$ & 4 & 32 & 1{,}125 & 64 & 19{,}683 \\
$C_4$ & 4 & 32 & 693 & 64 & 8{,}019 \\
paw & 4 & 32 & 765 & 64 & 8{,}019 \\
diamond & 4 & 32 & 549 & 64 & 4{,}131 \\
$K_4$ & 4 & 32 & 477 & 64 & 2{,}835 \\
$C_5$ & 5 & 64 & 3{,}249 & 128 & 60{,}507 \\
bull & 5 & 64 & 3{,}825 & 128 & 72{,}171 \\
\bottomrule
\end{tabular}
\end{table}

The simulator described below was validated against the generator \eqref{eq:generator} on $K_2$, $P_3$ and $K_3$ with $F=3$. From a fixed non-absorbing configuration on each graph we sampled the first transition $2\times10^5$ times. The total variation distance between the empirical and exact distributions of the next configuration was 0.0020 at most, of the order of the sampling error, and the mean waiting times (2.994, 0.998, 0.749) matched the exact values $1/\sum_{\mathrm{Act}}(o_{xy}/F)$ (3.000, 1.000, 0.750).

\subsection{Simulations}

The simulations below only illustrate the theorems. The process was simulated exactly with the Gillespie algorithm \cite{gillespie1977} restricted to active edges: an active edge is selected with probability proportional to $o_{xy}/F$, the total rate $\sum_{\mathrm{Act}}o_{xy}/F$ sets the exponential waiting time, and the update follows \eqref{eq:generator}. Initial cultures were independent and uniform on $[q]^F$. We used nine connected graphs (Table~\ref{tab:graphs}) generated or loaded with NetworkX \cite{hagberg2008}: the Florentine families \cite{padgett1993}, Zachary's karate club \cite{zachary1977}, the Les Mis\'erables co-appearance network \cite{knuth1993}, an Erd\H{o}s--R\'enyi graph $G(100,0.06)$, the ring lattice with $k=6$ \cite{watts1998}, a random $6$-regular graph, a triangular lattice, a Holme--Kim graph with tunable clustering \cite{holme2002}, and a connected caveman graph of ten $10$-cliques \cite{watts1999}. For each graph and each $(F,q)\in\{(2,2),(3,6),(3,12)\}$ we ran $40$ independent realisations on the same fixed graph, $1080$ runs in total; run $j=0,\dots,39$ used the seed $1000+j$ for both the initial configuration and the dynamics, and the random graphs were generated once with fixed seeds.

\begin{table}[htbp]
\centering\small
\caption{The nine graphs: order $n$, size $m$, number of triangles $\Delta(G)$, open-wedge mass $\omega(G)$ and transitivity $3\Delta(G)/\sum_x\binom{d_x}{2}$.}\label{tab:graphs}
\begin{tabular}{lrrrrr}
\toprule
Graph & $n$ & $m$ & $\Delta(G)$ & $\omega(G)$ & transitivity\\
\midrule
Florentine & 15 & 20 & 3 & 38 & 0.191 \\
Karate & 34 & 78 & 45 & 393 & 0.256 \\
Les Mis\'erables & 77 & 254 & 467 & 1407 & 0.499 \\
Erd\H{o}s--R\'enyi & 100 & 312 & 35 & 1761 & 0.056 \\
Ring lattice & 100 & 300 & 300 & 600 & 0.600 \\
Random $6$-regular & 100 & 300 & 29 & 1413 & 0.058 \\
Triangular lattice & 99 & 258 & 160 & 672 & 0.417 \\
Holme--Kim & 100 & 289 & 160 & 2177 & 0.181 \\
Caveman & 100 & 450 & 1120 & 250 & 0.931 \\
\bottomrule
\end{tabular}
\end{table}

\begin{table}[htbp]
\centering\small
\caption{Structure of absorbing configurations, averaged over $40$ runs per row: share of runs ending in consensus, mean number of domains, boundary fraction $|\bd|/m$, shares of triangles with two and with three boundary edges, total number of triangles with exactly one boundary edge over all $40$ runs, mean ratio $\omega(G_I)/\omega(G)$, and the largest clique number $\mathrm{cl}_{\max}$ of the domain quotient $G/\Pdom$ over the $40$ runs.}\label{tab:structure}
\setlength{\tabcolsep}{3pt}\footnotesize
\begin{tabular}{llrrrrrrrr}
\toprule
Graph & $(F,q)$ & cons. & domains & $|\bd|/m$ & $\Delta_2/\Delta(G)$ & $\Delta_3/\Delta(G)$ & $\sum\Delta_1$ & $\frac{\omega(G_I)}{\omega(G)}$ & $\mathrm{cl}_{\max}$\\
\midrule
Florentine & $(2,2)$ & 0.33 & 2.5 & 0.15 & 0.22 & 0.000 & 0 & 0.73 & 2 \\
 & $(3,6)$ & 0.00 & 5.8 & 0.44 & 0.50 & 0.108 & 0 & 0.36 & 3 \\
 & $(3,12)$ & 0.00 & 10.4 & 0.75 & 0.53 & 0.375 & 0 & 0.07 & 4 \\
\midrule
Karate & $(2,2)$ & 0.28 & 3.4 & 0.17 & 0.25 & 0.000 & 0 & 0.72 & 2 \\
 & $(3,6)$ & 0.07 & 5.3 & 0.27 & 0.36 & 0.034 & 0 & 0.57 & 4 \\
 & $(3,12)$ & 0.00 & 15.1 & 0.65 & 0.58 & 0.248 & 0 & 0.13 & 5 \\
\midrule
Les Mis\'erables & $(2,2)$ & 0.38 & 5.9 & 0.14 & 0.19 & 0.000 & 0 & 0.73 & 2 \\
 & $(3,6)$ & 0.00 & 9.2 & 0.16 & 0.21 & 0.006 & 0 & 0.70 & 3 \\
 & $(3,12)$ & 0.00 & 23.4 & 0.46 & 0.46 & 0.136 & 0 & 0.34 & 8 \\
\midrule
Erd\H{o}s--R\'enyi & $(2,2)$ & 0.35 & 2.6 & 0.20 & 0.30 & 0.000 & 0 & 0.67 & 2 \\
 & $(3,6)$ & 0.33 & 2.1 & 0.04 & 0.05 & 0.001 & 0 & 0.94 & 3 \\
 & $(3,12)$ & 0.00 & 10.8 & 0.37 & 0.42 & 0.082 & 0 & 0.46 & 6 \\
\midrule
Ring lattice & $(2,2)$ & 0.33 & 2.9 & 0.07 & 0.10 & 0.000 & 0 & 0.87 & 2 \\
 & $(3,6)$ & 0.00 & 12.2 & 0.30 & 0.39 & 0.042 & 0 & 0.53 & 4 \\
 & $(3,12)$ & 0.00 & 41.0 & 0.70 & 0.53 & 0.345 & 0 & 0.11 & 4 \\
\midrule
Random $6$-regular & $(2,2)$ & 0.25 & 2.3 & 0.19 & 0.27 & 0.000 & 0 & 0.69 & 2 \\
 & $(3,6)$ & 0.60 & 1.6 & 0.02 & 0.03 & 0.000 & 0 & 0.96 & 2 \\
 & $(3,12)$ & 0.00 & 7.5 & 0.33 & 0.39 & 0.058 & 0 & 0.51 & 6 \\
\midrule
Triangular lattice & $(2,2)$ & 0.45 & 2.2 & 0.08 & 0.12 & 0.000 & 0 & 0.85 & 2 \\
 & $(3,6)$ & 0.05 & 6.1 & 0.17 & 0.23 & 0.014 & 0 & 0.72 & 4 \\
 & $(3,12)$ & 0.00 & 33.6 & 0.61 & 0.53 & 0.253 & 0 & 0.19 & 4 \\
\midrule
Holme--Kim & $(2,2)$ & 0.23 & 4.8 & 0.22 & 0.34 & 0.000 & 0 & 0.63 & 2 \\
 & $(3,6)$ & 0.23 & 3.2 & 0.05 & 0.06 & 0.002 & 0 & 0.92 & 3 \\
 & $(3,12)$ & 0.00 & 14.8 & 0.32 & 0.37 & 0.065 & 0 & 0.52 & 5 \\
\midrule
Caveman & $(2,2)$ & 0.28 & 4.0 & 0.09 & 0.13 & 0.000 & 0 & 0.81 & 2 \\
 & $(3,6)$ & 0.00 & 16.1 & 0.37 & 0.45 & 0.067 & 0 & 0.39 & 5 \\
 & $(3,12)$ & 0.00 & 38.0 & 0.71 & 0.52 & 0.357 & 0 & 0.10 & 7 \\
\bottomrule
\end{tabular}
\end{table}

\begin{table}[htbp]
\centering\small
\caption{Assortativity and absorption time. The coefficients are averaged over the non-consensus runs, whose number is given in the third column; $\bar r_{\mathrm{feat}}$ is the mean of $r_f$ over features. The last two columns give the mean absorption time and the mean of the bound $\tfrac F2(Fn^2-Q(\sigma_0))$ of Corollary~\ref{cor:time}.}\label{tab:assort}
\begin{tabular}{llrrrrrr}
\toprule
Graph & $(F,q)$ & runs & $\bar r_{\mathrm{feat}}$ & $\rcul$ & $\rdom$ & $\E\tau$ & bound\\
\midrule
Florentine & $(2,2)$ & 27 & 0.17 & 0.17 & 0.22 & 23.5 & 212 \\
 & $(3,6)$ & 40 & 0.14 & 0.18 & 0.18 & 34.4 & 789 \\
 & $(3,12)$ & 40 & 0.02 & 0.07 & 0.07 & 15.9 & 866 \\
\midrule
Karate & $(2,2)$ & 29 & 0.14 & 0.14 & 0.20 & 26.2 & 1{,}122 \\
 & $(3,6)$ & 37 & 0.16 & 0.18 & 0.18 & 59.3 & 4{,}201 \\
 & $(3,12)$ & 40 & 0.09 & 0.12 & 0.12 & 41.1 & 4{,}622 \\
\midrule
Les Mis\'erables & $(2,2)$ & 25 & 0.27 & 0.27 & 0.32 & 64.9 & 5{,}840 \\
 & $(3,6)$ & 40 & 0.35 & 0.36 & 0.36 & 105.6 & 21{,}948 \\
 & $(3,12)$ & 40 & 0.19 & 0.21 & 0.21 & 94.0 & 24{,}152 \\
\midrule
Erd\H{o}s--R\'enyi & $(2,2)$ & 26 & 0.17 & 0.17 & 0.18 & 47.9 & 9{,}912 \\
 & $(3,6)$ & 27 & 0.11 & 0.11 & 0.11 & 150.4 & 37{,}123 \\
 & $(3,12)$ & 40 & 0.17 & 0.17 & 0.17 & 161.9 & 40{,}848 \\
\midrule
Ring lattice & $(2,2)$ & 27 & 0.74 & 0.74 & 0.80 & 169.4 & 9{,}912 \\
 & $(3,6)$ & 40 & 0.48 & 0.53 & 0.54 & 243.3 & 37{,}123 \\
 & $(3,12)$ & 40 & 0.20 & 0.26 & 0.26 & 78.5 & 40{,}848 \\
\midrule
Random $6$-regular & $(2,2)$ & 30 & 0.21 & 0.21 & 0.22 & 46.2 & 9{,}912 \\
 & $(3,6)$ & 16 & 0.13 & 0.13 & 0.13 & 151.0 & 37{,}123 \\
 & $(3,12)$ & 40 & 0.19 & 0.19 & 0.19 & 178.8 & 40{,}848 \\
\midrule
Triangular lattice & $(2,2)$ & 22 & 0.59 & 0.59 & 0.62 & 106.4 & 9{,}713 \\
 & $(3,6)$ & 38 & 0.47 & 0.48 & 0.48 & 214.1 & 36{,}379 \\
 & $(3,12)$ & 40 & 0.24 & 0.28 & 0.28 & 177.5 & 40{,}030 \\
\midrule
Holme--Kim & $(2,2)$ & 31 & 0.18 & 0.18 & 0.22 & 54.4 & 9{,}912 \\
 & $(3,6)$ & 31 & 0.15 & 0.15 & 0.15 & 157.9 & 37{,}123 \\
 & $(3,12)$ & 40 & 0.18 & 0.18 & 0.18 & 172.5 & 40{,}848 \\
\midrule
Caveman & $(2,2)$ & 29 & 0.67 & 0.67 & 0.74 & 227.8 & 9{,}912 \\
 & $(3,6)$ & 40 & 0.50 & 0.57 & 0.57 & 150.8 & 37{,}123 \\
 & $(3,12)$ & 40 & 0.20 & 0.26 & 0.26 & 54.8 & 40{,}848 \\
\bottomrule
\end{tabular}
\end{table}

Tables~\ref{tab:structure} and~\ref{tab:assort} summarise the absorbing configurations. In all $1080$ runs no triangle carried exactly one boundary edge (Theorem~\ref{thm:triadic}), no triangle carried three boundary edges when $q=2$ (Corollary~\ref{cor:cliques}), and the ordering $r_f\le\rcul\le\rdom$ held for every feature (Theorem~\ref{thm:chain}). In every absorbed configuration, moreover, adjacent blocks of the domain quotient carried different values of each feature $f$. This is the constructive content of Theorem~\ref{thm:colour}, each feature supplying a proper $q$-colouring of the quotient, and it implies in particular that the clique number of the quotient never exceeded $q$, as Table~\ref{tab:structure} confirms. With $q=2$ the strict inequality $\rcul<\rdom$ is common, because disconnected domains often repeat a culture; with $q=12$ cultures are rarely repeated and $\rcul$ and $\rdom$ coincide to two decimals, while $\bar r_{\mathrm{feat}}$ is smaller on six of the nine graphs and equal to $\rcul$ to two decimals on the Erd\H{o}s--R\'enyi, random regular and Holme--Kim graphs. For $q=2$, connectedness forces all cultures into a complementary pair $\{c,\bar c\}$ (proof of Corollary~\ref{cor:q2}), so $\Pi_f=\Pcul$ and $r_f=\rcul$, as the table shows. On six of the nine graphs the ratio $\omega(G_I)/\omega(G)$ decreases from $(F,q)=(2,2)$ to $(3,12)$ as the boundary grows; on the Erd\H{o}s--R\'enyi, random regular and Holme--Kim graphs it is largest at $(3,6)$, where consensus is frequent. On the caveman graph, where $170$ of the $250$ open wedges use an edge joining two caves, it falls to $0.10$ at $q=12$. The mean absorption times lie below the bound of Corollary~\ref{cor:time} by factors from about $9$ to about $750$.

\begin{figure}[htbp]
\centering
\begin{subfigure}[b]{0.52\textwidth}\centering
\includegraphics[width=\textwidth]{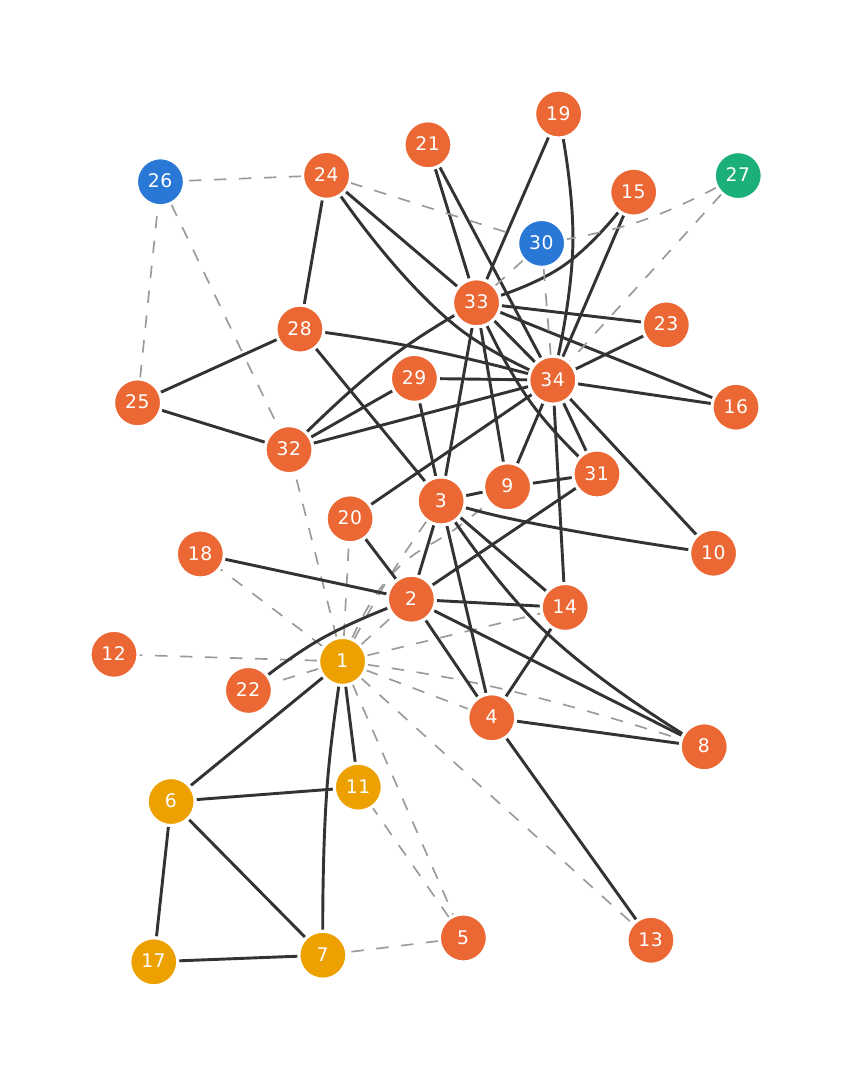}
\caption{Absorbing configuration}\label{fig:karate-a}
\end{subfigure}\hfill
\begin{subfigure}[b]{0.46\textwidth}\centering
\includegraphics[width=\textwidth]{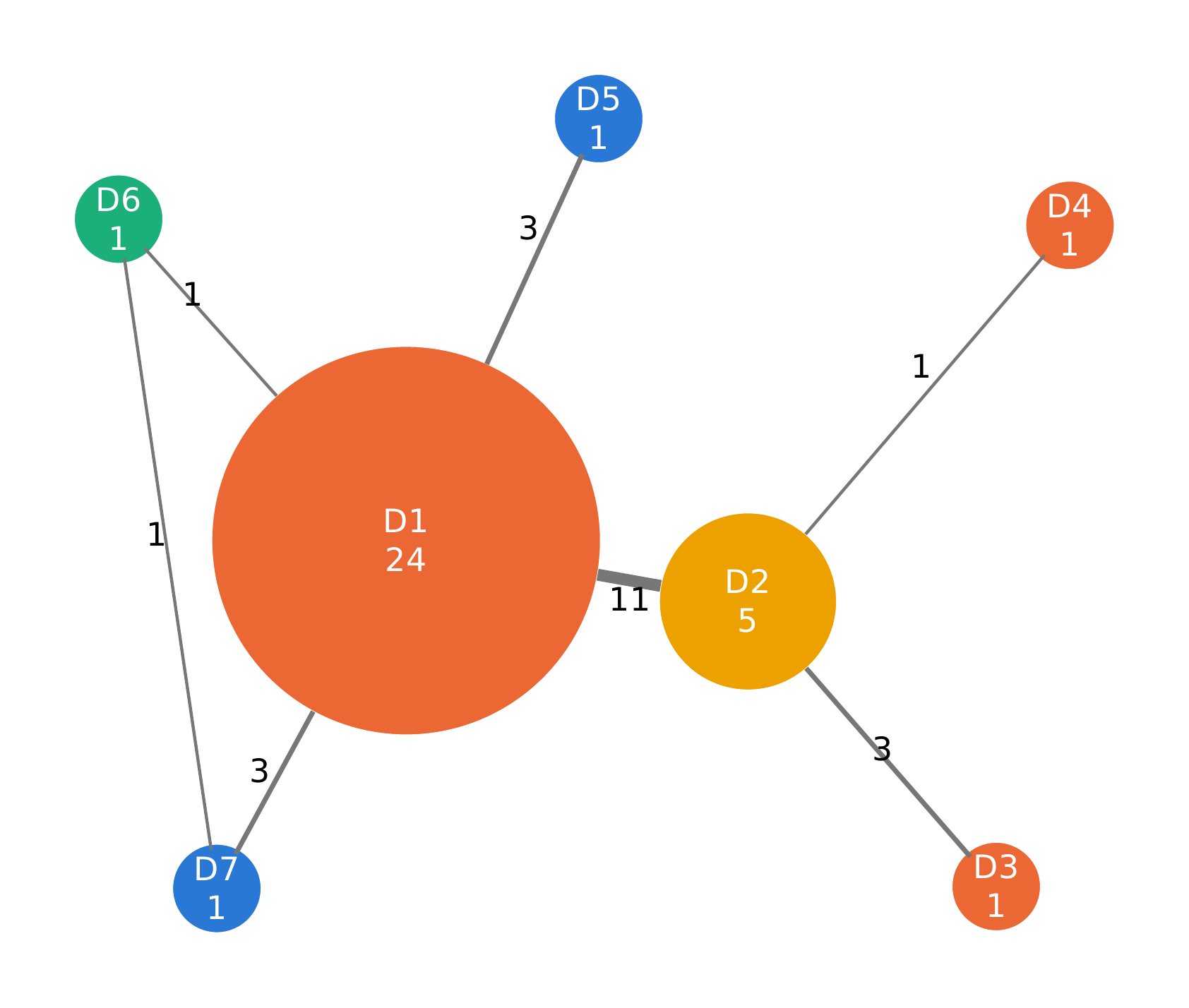}
\caption{Domain quotient $G/\Pdom$}\label{fig:karate-b}
\end{subfigure}
\caption{An absorbing configuration of the Axelrod model on the karate club with $F=3$, $q=6$. (a) Vertices are coloured by culture; solid edges are interior and dashed edges are boundary. (b) The domain quotient; node area is proportional to the number of vertices in the domain, colours are cultures, and edge labels count boundary edges. The weighted attribute assortativity of the quotient (self-loops weighted by interior edges) with respect to culture is $0.246$, equal to $\rcul$ for the configuration in (a).}\label{fig:karate}
\end{figure}

Figure~\ref{fig:karate} shows one absorbing configuration on the karate club with $F=3$ and $q=6$. It has seven domains but only four cultures: the large domain $D_1$ (24 vertices) shares its culture with the singletons $D_3$ and $D_4$, and $D_5$ shares its culture with $D_7$. Of the $78$ edges, $23$ are boundary edges. Of the $45$ triangles, $24$ are interior, $20$ have two boundary edges and one, formed by $D_1$, $D_6$ and $D_7$, has three; it appears as the triangle of the quotient and is admissible because $q\ge3$. The three feature coefficients are $0.237$, $0.237$ and $0.246$, while $\rcul=0.246$ and $\rdom=0.314$. The gap between $\rcul$ and $\rdom$ comes entirely from the repeated cultures, as Theorem~\ref{thm:chain}(iii) requires. The open-wedge mass drops from $\omega(G)=393$ to $\omega(G_I)=223$.

\begin{figure}[htbp]
\centering
\includegraphics[width=\textwidth]{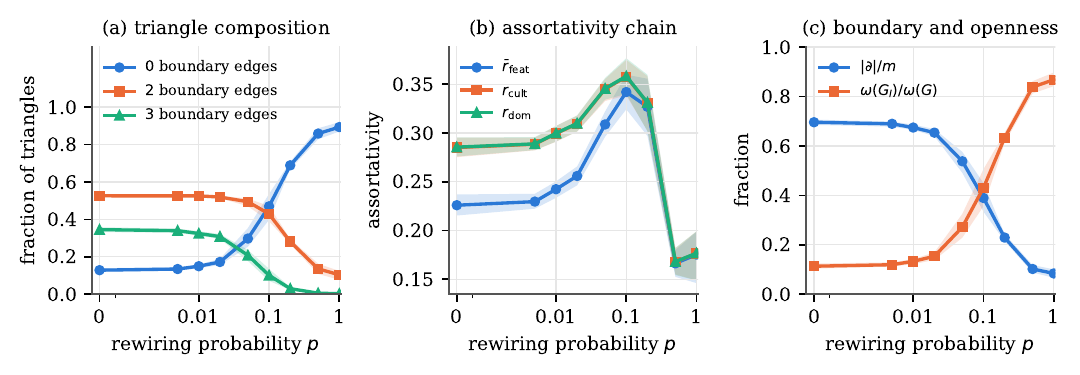}
\caption{Absorbing configurations on Watts--Strogatz graphs ($n=200$, $k=6$, $F=3$, $q=12$; $80$ runs per value of $p$). (a) Shares of triangles with $0$, $2$ and $3$ boundary edges; the share with exactly one boundary edge is identically zero. (b) Mean feature, culture and domain assortativity over non-consensus runs. (c) Boundary fraction and the ratio $\omega(G_I)/\omega(G)$. Bands are $95\%$ percentile intervals from a cluster bootstrap over the eight graph realisations ($2000$ resamples).}\label{fig:sweep}
\end{figure}

To see how triadicity of the host graph shapes the absorbing states, we rewired ring lattices with $n=200$ and $k=6$ by the Watts--Strogatz procedure \cite{watts1998}, using eight graph realisations (seeds $100,\dots,107$) and ten runs on each for every rewiring probability $p$, with $F=3$ and $q=12$. As $p$ grows from $0$ to $1$ the transitivity falls from $0.60$ to $0.02$. Figure~\ref{fig:sweep} reports the composition of the triangles, the three assortativity coefficients and the boundary and openness fractions. On the highly clustered lattices about seventy per cent of the edges are boundary edges and more than a third of the triangles have three boundary edges. As shortcuts destroy triangles and shorten paths, the boundary fraction falls below ten per cent, close to ninety per cent of the surviving triangles are interior, and $\omega(G_I)/\omega(G)$ rises towards one. No triangle has exactly one boundary edge (Theorem~\ref{thm:triadic}). The assortativity coefficients rise and then fall, and the chain $r_f\le\rcul\le\rdom$ is preserved throughout. We do not attempt to explain the location of the maximum, which presumably reflects the interplay between the number of domains and the concentration $s$.

\section{Discussion}\label{sec:discussion}

The main result of the paper is a classification of the absorbing configurations of the Axelrod model on a finite graph. A partition into connected blocks is a domain partition exactly when its quotient is properly $q$-colourable, the absorbing configurations with that domain partition are the homomorphisms from the quotient into $K_q^{\times F}$, and their total number is a sum of chromatic polynomials over connected partitions (Theorem~\ref{thm:colour}, Corollaries~\ref{cor:count} and~\ref{cor:q2}). The remaining results describe these configurations from the viewpoint of local graph structure. The interior graph is a disjoint union of induced subgraphs (Lemma~\ref{lem:induced}), triangles never carry exactly one boundary edge (Theorem~\ref{thm:triadic}), and the assortativity coefficients at the levels of domains, cultures and features share a common interior fraction and are ordered (Theorem~\ref{thm:chain}). On the dynamical side, the feature counts are martingales and the sums of their squares submartingales, with drift equal to $2/F$ times the total overlap on active edges (Theorem~\ref{thm:lyapunov}), which gives an absorption-time bound of order $F^2n^2$ and, on cliques, a lower bound on the concentration of final domain sizes.

Several questions remain open. The enumeration of Corollary~\ref{cor:count} counts absorbing configurations but says nothing about the probability with which the dynamics select them; apart from the degenerate case of Proposition~\ref{prop:selection}, describing the law of $\Pdom(\sigma_\tau)$ over the connected partitions, even for small graphs or for trees, is open. A second question concerns the dependence of the boundary fraction on the triadic structure of the host graph. Inequality~\eqref{eq:protection} says that edges sitting in many triangles with high-overlap legs cannot be boundary edges at a given time, and Figure~\ref{fig:sweep} suggests that this local protection alone does not decide the final boundary fraction, since the most clustered lattices freeze with the largest boundaries. A theorem relating $\E|\bd(\sigma_\tau)|$ to the triangle multiplicities $t_e$ or to the open-wedge mass $\omega(G)$, even for restricted families such as circulant graphs, is lacking. A third question concerns the vertex-driven variant of the model, in which each vertex updates at rate one by consulting a uniformly chosen neighbour. There the degree-weighted counts $\sum_{x:\sigma_f(x)=a}d_x$ are martingales, as for the voter model, and the degree-weighted concentrations $s(\Pi_f)$ in Theorem~\ref{thm:chain} would play the role that $Q_f/n^2$ plays here. A fourth concerns higher-order interactions: replacing pairwise edges by hyperedges along which cultures are compared would lead to analogues of the operators of \cite{boudourides_jcn} for hypergraphs. Finally, none of the present arguments addresses the infinite-volume questions of clustering versus fixation, which for graphs with triangles, such as the triangular lattice, remain open, as far as we know.

\section*{Data and code availability}
All computations in Section~\ref{sec:numerics} use public or synthetic graphs. The code that produces the simulations, the exhaustive enumeration, the tables and the figures, together with the numerical outputs, is available at \url{https://github.com/mboudour/axelrod-absorbing-states}.

\section*{Declarations}
\noindent\textbf{Funding.} This work received no specific grant from any funding agency, commercial or not-for-profit sectors.\\
\textbf{Competing interests.} The author declares no competing interests.\\
\textbf{Use of generative AI.} During the preparation of this work the author used ChatGPT (OpenAI) to assist with language editing, the correction of grammatical errors, and with checking bibliographic details. After using this tool, the author reviewed and edited the content as needed and takes full responsibility for the content of the published article.

\bibliographystyle{amsplain}
\bibliography{references}

\end{document}